# A Multilingual Python Programming Language

Saad Ahmed Bazaz
*Software Engineering and Product Design*
Grayhat Developers PVT Ltd
Islamabad, Pakistan
bazaz@grayhat.studio

Mirza Omer Beg
*Department of Artificial Intelligence*
National University of Computer and Emerging Sciences
Islamabad, Pakistan
omer.beg@nu.edu.pk

*Abstract*—All widely used and useful programming languages have a common problem. They restrict entry on the basis of knowledge of the English language. The lack of knowledge of English poses a major hurdle to many newcomers who do not have the resources, in terms of time and money, to learn the English language. Studies show that people learn better in their own language. Therefore, we propose a language transpiler built on top of the Python programming language, called UniversalPython, which allows one to write Python in their own human language. We demonstrate the ability to create an "Urdu Python" with this transpiler. In the future, we aim to scale the language to encapsulate more human languages to increase the availability of programming. The source code for this transpiler is open-source, and available at https://github.com/universalpython/universalpython

## I. Introduction

In our interconnected world, English often serves as the dominant language in technology and communication. However, a majority of the world still struggles with it [1], which creates significant barriers to education, particularly in fields like computer science and programming.

As coding education grows through boot camps and online platforms, it's crucial to embrace linguistic diversity. Research demonstrates that incorporating localized languages into curricula enhances students' comprehension and engagement in learning activities. This need is particularly pressing in developing regions, where there is an urgent call to democratize access to technology education. By aligning educational content with students' native languages, we create a more inclusive environment that encourages diverse participation in tech [2].

Our research seeks to challenge existing linguistic barriers by proposing a multilingual wrapper on the Python programming language, specifically designed for non-English speakers. By enabling coding in one's native human language, in a fairly popular programming language like Python, we empower a broader demographic of learners to engage with technology without the constraints of language proficiency. This approach has the potential to reshape how coding is taught and learned, fostering a more equitable future in tech education.

## II. Related Work

Since the inception of high-level computer programming languages, English has been the predominant language used. FORTRAN, the first widely adopted high-level programming language, utilized an English instruction set [3]. Interestingly, however, the first high-level programming language ever recorded, "Plankalkül," was developed by the German engineer Konrad Zuse [4]. This language is often regarded as a forerunner to modern programming languages [5].

### A. Learning in your own language

Research indicates that students in K-12 education tend to learn more effectively in their localized language [6]. Various successful initiatives around the globe have aimed to integrate non-dominant languages into school curricula, fostering inclusivity, preserving culture, and enhancing comprehension [7]. In several developing regions, language can act as a significant barrier to education, leading youths in rural areas to drop out when instruction is delivered in a non-native language [8]. Literacy classes conducted in students' mother tongues have proven beneficial in re-integrating these young individuals into formal education. Notably, girls in particular have demonstrated marked improvements in achievement, self-image as learners, and retention rates when educated in their local language [9].

Computing education is particularly challenging, due to the overarching presence of English in programming. Efforts to localize computing education content have shown promise, as exemplified by a Qatari middle school curriculum [10]. Longitudinal studies spanning five countries and involving over 15,000 users of Scratch, a prominent informal learning platform, reveal that novice users coding with localized programming language keywords and environments grasp new programming concepts more rapidly compared to those using English [11]. This suggests potential pathways to "universal design" [12] by internationalizing code examples, making programming more accessible and robust for diverse learners. Furthermore, an exciting parallel emerges: flipping the "language bit" may also facilitate English speakers in learning foreign languages within their usual programming practices [13].

## B. Non-English, monolingual programming languages

The history of localized educational programming languages began as early as 1975 during the USSR era with the development of Robic, a Russian programming language designed for children aged 8 to 11 [14]. This language was later adapted into a software system called "Schoolgirl" for the Agat computer, utilizing syntax derived from the Russian vocabulary.

Since then, various programming languages have emerged, focusing on single languages (monolingual) distinct from English. For instance, a Yoruba-based programming language [15] and a Hindi programming language named Kalaam [16] exemplify this trend. However, many of these initiatives fall short due to a lack of community engagement and interest, hindering their practical applicability beyond simply teaching programming constructs. Consequently, these languages often become "another syntax to learn."

In a more recent development, a programming language in Ancient Chinese [17] took the internet by storm, inspiring hundreds of thousands of users to create engaging programs. Additionally, Alif, an Arabic programming language with a Pythonic syntax [18], has emerged. Alif is crafted in C/C++ and aligns itself with Python, allowing programmers to explore it for enjoyment and educational purposes, though it faces challenges in keeping pace with Python's continued advancements.

## C. Multilingual programming languages

There have been several attempts to create multilingual programming languages, designed to simplify programming concepts for novices through the use of symbols and illustrations [19]. These languages often label the symbols, localizing them to cater to specific audiences. Most of these languages serve primarily educational roles.

Notably, Hedy [20] stands out as a text-focused programming language and environment tailored for teaching beginners how to code. Scratch, a well-known example in this domain, has also played a pivotal role in engaging learners through its visual and interactive programming approach [21].

Such attempts are great for learning, yet require a large amount of funding for the maintenance of the entire ecosystem. Also, these tools and programming languages are mostly restricted to basic operations. However, some content creators like *Griffpatch* on YouTube [22] have been able to produce impressive games with tools like Scratch.

A subsequent consequence of designing a programming language from the ground up which supports all human languages, can possibly lead to a "Turing tarpit" problem [23], which refers to a programming language which is so flexible, almost nothing worth building can be built in it, without a huge amount of complexity. Any programs with slightly heavier-than-usual computing requirements, or with multiple transformations, can become cumbersome in such a language.

## D. Localizing Existing Programming Languages

PseuToPy [24] introduces an intermediate pseudo language designed to interact with the Python interpreter. This method can complicate the user experience due to possible inconsistencies in machine translation. Furthermore, PseuToPy lacks a standardized structure, complicating transitions between languages.

UrduScript [25], a JavaScript dialect in Urdu, aims to "... make programming more accessible to beginners from South Asia" [26].

Another notable project is Chinese Python [27], an open-source initiative that adapts Python source code to incorporate Chinese symbols. However, applying this to various human languages would necessitate a complete recompilation of the Python programming language for each language and nearly every major version.

While these projects employ familiar syntax due to their foundations in existing languages, maintaining them across new versions and scaling them to different human languages remains challenging.

## E. Research Gap

As outlined earlier, localized programming languages face several key challenges:

- Community adoption issues stemming from unfamiliar syntax and rules
- Maintenance difficulties arising from complexity or high overhead
- Limited scalability to other human languages
- Visual tools and programming languages primarily supporting only basic operations
- Insufficient third-party library support

The difficulties encountered by both monolingual and multilingual programming languages are well summarized in [28], which presents a framework for evaluating the localization of a programming language. Given these challenges, we pose the following questions:

- **Can a programming language designed to facilitate education in localized languages also be easy to maintain, have a low learning curve, and support novices in advancing to professional levels?**
- **Can this language be flexible enough to facilitate the development of complex programs and maintain interoperability with existing libraries and languages?**

Today, Python welcomes a broader range of Unicode, allowing variable names in various languages to be both interpreted correctly and referenced accurately [29].

The Legesher [30] project aims to empower individuals to code in any programming language using their native language. Currently in beta, it holds promise by enabling the community to define translations of programming languages through simple YAML files, supporting major scripting languages like Python, JavaScript, and Julia.

Universal Python [31], another relevant project, proposes a straightforward solution: a transpiler built over the Python language, allowing users to code in their native languages (e.g., French Python, Italian Python, etc.).

To solve our research gap, we build on top of the Python language itself. We not only implement the idea in [31], but also benchmark it against Python and other languages. Finally, we abstracting it for broader application across other human languages.

### III. Design of UniversalPython

We have implemented a simple framework, illustrated in Figure 1, which serves as a transpiler (also known as a **source-to-source compiler**) on top of the Python interpreter. This transpiler translates a higher-level, non-English variant of the Python programming language (e.g., Urdu Python, Chinese Python, Hindi Python, Arabic Python, etc.) into the standard Python programming language, which is in English.

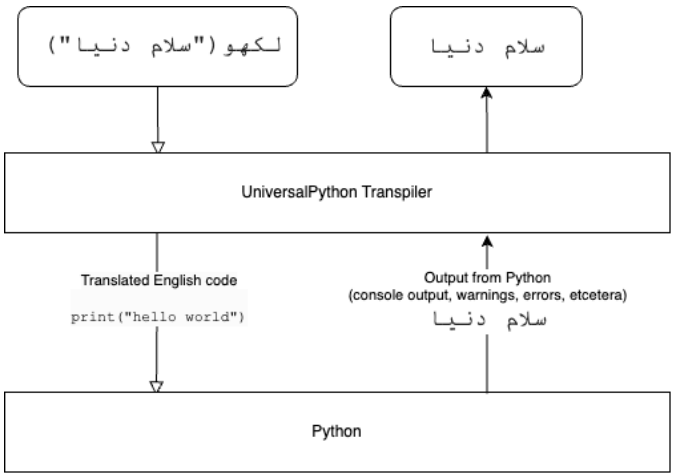

Figure 1: A high-level abstraction of how the UniversalPython transpiler works.

#### A. Walkthrough

For instance, if a user writes the following code in Urdu[1]:

کچھ=۲
اگر کچھ==۱:
لکھو("سلام دنیا")
ورنہ:
لکھو("خدا حافظ")

This code is passed to "UniversalPython," which first translates it to English using Lexical Analysis and Parsing with the PLY library. To achieve this, it initially loads the Urdu dictionary, a YAML file containing mappings from Urdu to English. This dictionary maps each Urdu word to its corresponding Python keyword. An example of such a dictionary is provided in Table I.

[1]Throughout this paper, we take the example of *Urdu* (اردو), the national language of Pakistan (the author's home country), and a traditional right-to-left language, as an example translation language.

In PLY, we can reserve certain keywords, allowing the library to automatically tokenize them. We establish a Grammar Rule whereby, whenever a reserved keyword (i.e., a word present as a key in the language dictionary) is tokenized, the lexer simply searches for that token in the language dictionary (key) and replaces it with the corresponding English keyword (value). Additionally, we set a Grammar Rule to ignore all content within double or single quotes (i.e., strings and docstrings) and content within comments (which begin with a #). We accomplish this using Regular Expressions.

Urdu numbers also exist within the Unicode scale. For instance, ۰ (or roman 0) is mapped to 1776, while ۹ (or roman 9) is at 1785.

With this in mind, we define a Regular Expression that detects any symbols within the range of 1776 to 1785, which corresponds to Urdu digits. We then utilize the same language dictionary to translate these numbers into Roman numerals. For example, ۵ becomes 5, ۹۰ becomes 90, ۱۰ translates to 10, and ۲۰۲۵ becomes 2025. Furthermore, we replace all periods (.) and commas (,), as these appear differently in Urdu compared to English.

The remaining code is left untouched. Symbols such as :, ;, etc., or any lexer errors are ignored to preserve the original structure of the code as much as possible. There is no need to translate such symbols and errors; they are meant to be handled by Python, not by UniversalPython. Referring back to our initial example code, our transpiler detects and replaces it with 'print', translates اگر to 'if', replaces ورنہ with 'else', and changes all Arabic digits to Roman numerals. Consequently, the translated code reads as follows:

```
khchh = 2
if khchh == 1:
 print ("سلام دنیا")
else:
 print ("خدا حافظ")
```

The above code is essentially vanilla Python code that can now be executed by the Python interpreter. The Urdu variable name is automatically converted to English using the unidecode library. However, even if it were in Arabic script, this would not pose an issue for newer versions of Python, as support for Unicode variable names has been available since Python 3.0 [29].

Returning to the example, this code is then passed to the Python interpreter, which outputs a response. This response could include print statements entered by the user, compiler/interpreter warnings, or errors. This output is then relayed back to UniversalPython, where it is tokenized again to replace keywords in the event of error messages. In our example, since there are no errors, it simply outputs the response as-is. Thus, the response would be:

سلام دنیا

Table I: Language dictionary for periods and commas

| Key | Value |
|-----|-------|
| .   | ۔     |
| ,   | ،     |

## B. Language dictionary

The language dictionary for the Urdu variant of UniversalPython is a YAML file containing mappings from Urdu to English keywords. The PLY library enables the reservation of keywords for automatic tokenization and replacement with their English equivalents. Additionally, grammar rules are established to ignore content within quotes, comments, and Urdu numbers (which exist on the Unicode scale).

Table II: Comparison of Python Keywords and their Urdu Equivalents

| Python (original) | Urdu (اردو) |
|-------------------|-------------|
| 'print'           | لکھو        |
| 'if'              | اگر         |
| 'elif'            | ورنہ اگر    |
| 'else'            | ورنہ        |
| 'while'           | جبتک        |
| 'for'             | جو          |
| 'in'              | اندر        |
| 'input'           | داخلہ       |
| 'break'           | توڑ         |
| 'continue'        | جاری        |
| 'pass'            | گزر         |
| 'True'            | حق          |
| 'False'           | باطل        |

## IV. EXPERIMENTATION

### A. Evaluation metrics

We propose the following metrics which UniversalPython should meet, so it can be considered an effective solution for our research problem:

1) Programs which work in Python, should work in UniversalPython.
2) UniversalPython should operate as a reasonable speed which at least does not disturb the programmer.
3) UniversalPython should be able to translate from one non-English language to another
4) A comparison should be made against other existing non-English, monolingual programming languages and multilingual programming languages
5) A user experience test should be conducted to find out user acceptability towards a language in their native tongue.

### B. Benchmarks with Python

**Time:** Check the execution time/performance of Python vs UniversalPython. We use a benchmarking tool called hyperfine which runs each program multiple times to produce a mean execution time.

**Conversion:** Convert simple programs from English Python to UniversalPython, and test if they still work. Our testing mechanism for the above is as below:

1) We take an existing Python program, lets say multiplication.py.
2) Run the UniversalPython system in reverse. This will flip the language dictionary and generate an Urdu version of the program (all the keywords would be translated from English to Urdu). Save it as multiplication.ur.py.
3) Run multiplication.ur.py using UniversalPython, and multiplication.py using Python.

If both give the same output on the terminal then no data loss has occurred; Both Urdu program and English program output the same message. Hence we can say that the program has safely converted from English to UniversalPython and vice versa without breaking the code or changing the logic.

We take simple algorithmic programs from TheAlgorithms/Python [32], a repository containing implementations of well-known algorithms, in the Python programming language. We run a loop over them and run the above algorithm on each to test.

## V. RESULTS

### A. Benchmarking Python and UniversalPython Code

We conducted our benchmarking experiments on a MacBook Air (Late 2019).

**Execution Success Rate:** Table IV describes our results over 110 simple Python programs present in TheAlgorithms/Python [14] math implementation.

We found that 98% of simple programs could be automatically converted into Urdu Python without disrupting the code. The two programs that failed encountered an issue with generating the Urdu version of an English program (i.e., running UniversalPython in reverse). The failure stemmed from the inability to distinguish between "is" and "==," as they serve the same purpose in English Python.

**Execution Time:** Upon summarizing the experiment, we concluded that UniversalPython performed better with simpler and less wordy programs. In contrast, for more verbose and lengthy algorithms, Python showed better performance in terms of execution time. Nevertheless, both languages operated within an acceptable speed range. Some notable performance differences among various algorithms are illustrated in Tables IV and VI.

### B. Exploring Different Languages and Interlanguage Translation

Our framework has the potential to be generic enough to span multiple languages. We successfully developed three variants—Urdu, Chinese, and Hindi—to showcase UniversalPython's capability to extend across various languages.

Table III: Comparing UniversalPython with other non-English, monolingual programming languages and multilingual programming languages using "A Framework for the Localization of Programming Languages" [28]

| Aspect/Prog Language | UniversalPython | Scratch | Wenyan | Yoruba | Kalaam | Alif | Chinese Python | Hedy | Legesher |
|---|---|---|---|---|---|---|---|---|---|
| Language | Multi | Multi | Chinese | Yoruba | Hindi | Arabic | Chinese | Multi | Multi |
| Alignment | N | N | N | N | N | T | N | NT | NT |
| Non-English keywords | ● | ● | ● | ● | ● | ● | ● | ● | ● |
| Non-Latin variable names | ● | ● | ● | ● | ● | ● | ● | ● | ● |
| Non-English productions | ● | ○ | ○ | - | - | ● | ○ | - | ● |
| Non-English numerals | ● | - | ● | - | - | ● | ● | ● | ● |
| Characters without meaning | ● | - | ● | - | - | ● | ● | ○ | ● |
| Diacritics | ● | ● | ○ | ○ | ○ | ● | ○ | ● | ● |
| Alternative keywords | ● | - | ● | ● | ○ | - | ● | ● | ● |
| Localized punctuation | ● | ○ | ● | ○ | ○ | ● | ● | ● | ● |
| Right to left support | ● | ○ | ○ | - | ○ | ● | ○ | ● | ○ |
| Multi-lingual programming | ● | - | - | - | - | - | - | ● | ● |
| Error messages | - | ○ | - | ● | ○ | ● | - | ● | ● |
| Multi-lingual 3rd-party libraries | ○ | ○ | ○ | - | ○ | ○ | ○ | ● | ● |

Table IV: Results from conversion test

| Execution Status | Number of Programs |
|---|---|
| PASS | 108 |
| FAIL | 2 |

Table V: Execution times of functions where our transpiler performed well.

| Function | Python | UniversalPython | Difference |
|---|---|---|---|
| softmax | 0.408194 | 0.061837 | 0.346357 |
| gaussian | 0.873841 | 0.558138 | 0.315703 |
| radix2 fft | 0.473321 | 0.187514 | 0.285806 |

Table VI: Execution times of functions where our transpiler performed poorly.

| Function | Python | UniversalPython | Difference |
|---|---|---|---|
| simpson rule | 0.168146 | 0.428384 | 0.260238 |
| quad eqs complex num | 0.162974 | 0.400886 | 0.237912 |
| square root | 0.165146 | 0.330606 | 0.16546 |

### C. Creating the Jupyter Kernel

To illustrate the ease with which plugins can be created for UniversalPython, we developed a wrapper for the IPython kernel. In this implementation, we imported UniversalPython as a package and translated the code (i.e., from Urdu to English) before passing it to IPython. By overriding the `do_execute` and `do_complete` functions in IPython, we achieved a functional kernel for UniversalPython that operates line by line while maintaining program memory. This provided us with a visual interface for comprehensive testing of the language.

### D. User Testing through a Hackathon

On May 21, 2022, we organized a "speed algorithm coding" hackathon at the National University of Computer and Emerging Sciences. During this event, we introduced UniversalPython at the outset, presenting it as the programming language for the challenge, but limited it to just an Urdu dictionary.

The participants, aged between 13 and 28 and currently enrolled in high school or equivalent programs, or pursuing a Bachelor's degree in Computer Science (or equivalent), had preliminary knowledge of coding in Python and was familiar with implementing algorithms.

The initial response to the new language was surprising; many participants were hesitant to use it. However, after consulting the documentation and understanding the syntax, more than 80% of the participants successfully submitted their coding challenges. We received mixed feedback, with some participants appreciating the translation of Python, particularly those from remote areas in Pakistan, whereas others felt more comfortable coding directly in Python.

Furthermore, we demonstrated language translation, for instance, from Urdu to Hindi (as depicted in Figure 2), using English as an *intermediate language.*

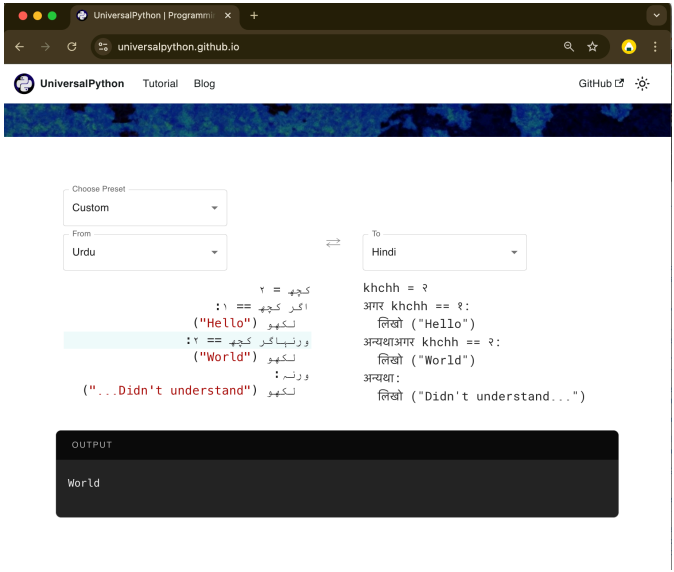

Figure 2: Urdu to Hindi code translation and successful compilation with UniversalPython.

## VI. LIMITATIONS

The growth and success of any software platform relies heavily on robust third-party library support. These libraries can allow developers to leverage existing solutions rather than building from scratch. However, in the context of UniversalPython, one notable limitation is the lack of translation support for these third-party libraries. Although these libraries may not directly lead to program crashes, it can significantly impact the overall developer experience. Developers often rely on documentation and error messages that are usually in English. Without translating these resources, non-English speakers may struggle to fully understand and integrate these libraries into their projects. A more seamless approach where third-party libraries and their associated documentation are fully translated would greatly improve accessibility and foster a more inclusive development environment.

The efficacy of translation tools can vary dramatically, particularly for low-resource languages or those with intricate contextual uses. While UniversalPython aims to bridge the gap between programming languages and everyday language, the quality of translations for certain languages may fall short, due to limitations of machine translation engines like Google Translate, leading to confusion or misinterpretation of code. Poor translation quality can be particularly detrimental when developers rely on translated keywords, function names, or error messages for debugging and learning. Some languages may lack the dataset or tools necessary for accurate translation, resulting in vague or inaccurate terms being utilized in the programming context. This limitation not only hinders the coding experience for non-native speakers but may also lead to errors in code execution.

Another challenge within UniversalPython is the inability to translate words or phrases that contain spaces. For example, the Python keyword 'elif' translates to ورنہ اگر in Urdu, which presents a challenge due to the space separating the components of the translation. This issue can pose limitations not just for keywords but also for function names and variable declarations, which may have varying syntactical structures across languages. Additionally, many languages differ in their grammatical order, such as the placement of verbs and nouns, which can further complicate translation efforts. Furthermore, some languages require different terms based on quantity, like using distinct forms for singular versus plural. In creating a more user-friendly experience for UniversalPython users, thoughtful consideration must be given to these linguistic nuances. A more flexible developer experience may help accommodate these differences and allow developers to write and comprehend code more naturally in their native language.

## VII. FUTURE WORK

To enhance the development of UniversalPython, we envision it as an extensible programming language that evolves alongside Python, ensuring seamless interoperability. Its maintainability is a significant advantage; since it functions as a simple wrapper, only the keywords in the dictionary need to be updated to accommodate any future or past Python versions. This would, however, entail a better versioning mechanism for the dictionaries to maintain parity with Python versions.

A key area for improvement lies in our goal to incorporate as many Python keywords as possible, ideally encompassing the entire set. Additionally, we should focus on translating major libraries and packages, such as pandas, numpy, and cv2. One efficient approach to achieve this could involve automatically scanning the source code of these libraries to identify critical keywords, translating them into their most suitable equivalents in the supported languages, and integrating them into our grammar. This process could leverage advanced technologies like Natural Language Processing, Deep Learning, and Machine Learning.

In drawing parallels with TypeScript's role in the JavaScript ecosystem [33], we can inspire UniversalPython's functionality as a build-time transpiler to its underlying Python code. To facilitate third-party library integration, we could learn from TypeScript's *\*.d.ts* files. In our case, this might take the form of an *interfaces.<language_code>.yml* file, outlining translations for functions and variables in a structured format. We leave the implementation details to the imagination of the reader.

Enhancing user experience is another vital aspect. We could introduce a feature allowing users to choose whether they want the system to detect variables and functions and translate them using machine translation into their corresponding names. Offering this level of customization could significantly enrich the user experience.

To ensure translation accuracy, we propose developing a human-involved translation verification system. This system could engage both beginner and experienced programmers by asking them to express how they would write a specific code snippet in their native language. Using data from platforms such as Leetcode for this purpose could yield valuable insights, and we leave this idea open for further exploration.

Conducting extensive User Experience tests with programmers from diverse backgrounds is crucial to determine if UniversalPython genuinely aids non-English speakers in their programming journey. Gathering feedback from such a wide demographic will help us refine our approach and ensure it meets the needs of its users.

Additionally, it is essential to design and develop a robust evaluation metric to assess the effectiveness of UniversalPython and similar multilingual programming languages or frameworks.

Our framework is designed to seamlessly support UniversalPython plugins within existing Python Integrated Development Environments (IDEs). By acting as a bridge, it translates non-English code into English before passing it on to the IDE. We can also develop a Read-Eval-Print Loop (REPL) by integrating it into the Python REPL, along with potential integrations with platforms like Replit and Leetcode.

Finally, securing official support from the Python Software Foundation or related organizations could significantly propel this project forward. Such endorsement would enhance credibility and provide a strong foundation for broader adoption of UniversalPython, ultimately benefiting non-English speaking programmers on a global scale.

## Acknowledgment

I would like to acknowledge my mentor, Dr Omer Beg, for always guiding me to the right path during my Bachelor's, professional life, and Master's degree. Without him, this project would have remained shelved for a few years further, since its inception in late 2021. I would also like to acknowledge my friend Syed Zohair Abbas Hadi for his technical support in creating the first draft of this paper.